\documentclass[
reprint,
superscriptaddress,
nofootinbib,
amsmath,amssymb,amsart,
aps,
pra,
]{revtex4-1}

\usepackage{graphicx}
\usepackage[caption=false]{subfig}
\usepackage{dcolumn}
\usepackage{bm}
\usepackage{color}
\usepackage{siunitx}
\usepackage{upgreek}

\newcommand{\bra}[1]{\langle #1|}
\newcommand{\ket}[1]{|#1\rangle}

\newcommand{\bfa}{\mathbf{a}}

\newcommand{\reviewcolor}{black}

\begin{document}

\title{Experimental implementation of a Raman-assisted six-quanta process}

\author{S.O.~Mundhada}
\email[Electronic address: ]{shantanu.mundhada@yale.edu}
\author{A.~Grimm}
\author{J.~Venkatraman}
\author{Z.K.~Minev}
\author{S.~Touzard}
\author{N.E.~Frattini}
\author{V.V.~Sivak}
\author{K.~Sliwa}
\altaffiliation{Present address: Quantum Circuits Inc., New Haven, CT 06511}
\author{P.~Reinhold}
\author{S.~Shankar}
\affiliation{Department of Applied Physics, Yale University, New Haven, CT 06511.}
\author{M.~Mirrahimi}
\affiliation{QUANTIC team, INRIA de Paris, 2 Rue Simone Iff, 75012 Paris, France}
\author{M.H.~Devoret}
\email[Electronic address: ]{michel.devoret@yale.edu}
\affiliation{Department of Applied Physics, Yale University, New Haven, CT 06511.}
\date{\today}
\begin{abstract}
\textcolor{\reviewcolor}{Nonlinear processes in the quantum regime are essential for many applications, such as quantum-limited amplification, measurement and control of quantum systems. In particular, the field of quantum error correction relies heavily on high-order nonlinear interactions between various modes of a quantum system. However,} the required order of nonlinearity is often not directly available or weak compared to dissipation present in the system. Here, we experimentally demonstrate a route to obtain higher-order nonlinearity by combining more easily available lower-order nonlinear processes, using a generalization of the Raman transition. In particular, we show a transformation of four photons of a high-Q superconducting resonator into two excitations of a superconducting transmon mode and vice versa. The resulting six-quanta process is obtained by cascading two fourth-order nonlinear processes through a virtual state. \textcolor{\reviewcolor}{We expect this type of process to become a key component of hardware efficient quantum error correction using continuous-variable error correction codes.}
\end{abstract}

\maketitle

\begin{figure}[t!]
\centering
\includegraphics[width=0.45\textwidth]{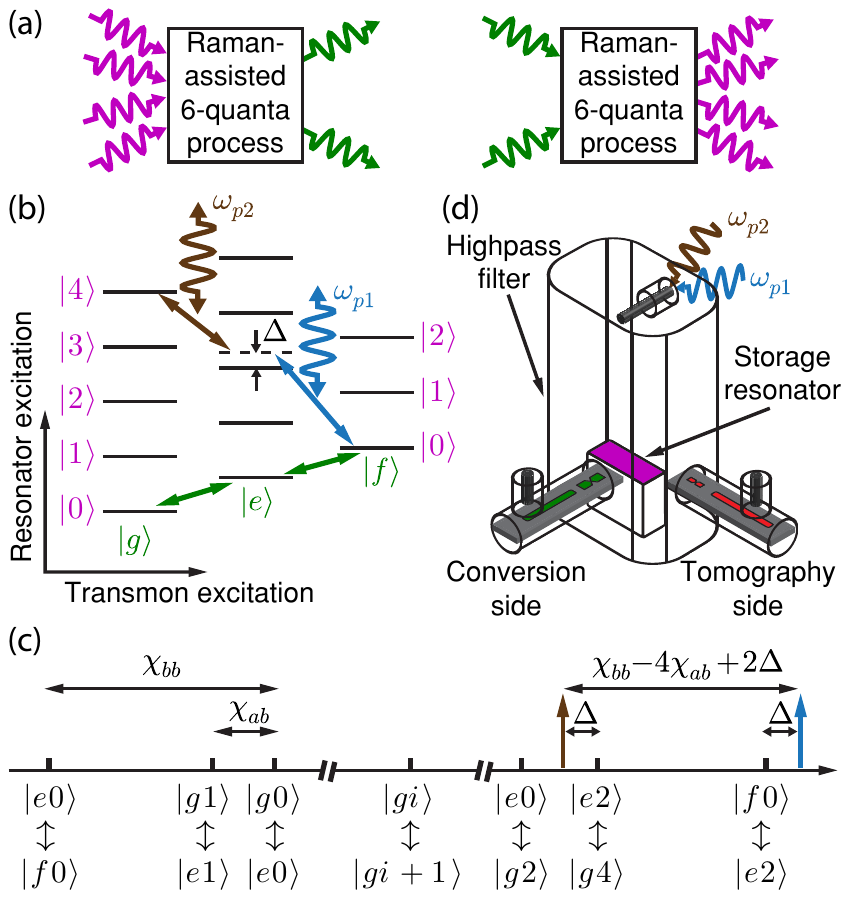}
\caption{\textbf{Schematic of Raman-assisted nonlinear processes and their experimental implementation.} (a) The target six-quanta process that exchanges four photons of a high-Q resonator (magenta) with two excitations of a transmon mode (green) and vice versa. (b) Energy level diagram of a high-Q storage resonator at frequency $\omega_a$ coupled to a transmon mode at frequency $\omega_b$ (called conversion mode). The first three transmon eigenstates (denoted by $\ket{g}$, $\ket{e}$ and $\ket{f}$) and the first five eigenstates of the storage resonator (denoted by $\ket{0}$ to $\ket{4}$) are considered. Starting in $\ket{g0}$, the system is prepared in $\ket{f0}$ by applying $\ket{g}\rightarrow\ket{e}$ and $\ket{e}\rightarrow\ket{f}$ Rabi pulses (green arrows). A pump at frequency $\omega_{p1}$ (blue) connects $\ket{f0}$  to a virtual \textcolor{\reviewcolor}{(non-energy-conserving)} state, represented by the dashed line, detuned from $\ket{e2}$ with a detuning $\Delta$. \textcolor{\reviewcolor}{This virtual state acts as an intermediate metastable excitation of the transmon.} A second pump at frequency $\omega_{p2}$ (brown) connects the virtual state to $\ket{g4}$, thus converting the two transmon excitations into four resonator excitations. (c) Frequencies of the pumps and the transitions involved in the scheme. (d) Schematic of the implementation. The high-Q storage mode is formed by an aluminum $\lambda/4$-type 3-dimensional superconducting resonator (magenta), which is dispersively coupled to the conversion transmon (green) and the tomography transmon (red). The two $\lambda/2$ stripline resonators coupled to the transmons are used for performing single-shot readout of the respective transmons. \label{fig:figure_1}}
\end{figure}
\section{Introduction}
\textcolor{\reviewcolor}{Encoding quantum information in the large Hilbert space of a harmonic oscillator allows for hardware-efficient quantum error correction~\cite{Gottesman2001,Lassen2010,Leghtas2013,Albert2016,Ofek2016}. A further increase in hardware efficiency can be achieved by protecting the information using an autonomous feedback mechanism. It is possible to achieve such autonomous quantum error correction by using nonlinear driven-dissipative processes to create a decoherence-free manifold of quantum states, within the Hilbert space of the oscillator~\cite{Wolinsky1988,Zanardi1997,Lidar1998,Kempe2001,Cohen2014,Mirrahimi2014,Albert2016a,Leghtas2015,Kapit2016,Kapit2017,Puri2017,Touzard2018,Albert2019}.} In particular, a stabilized manifold spanned by four coherent states of a harmonic oscillator has been proposed for the implementation of a hardware efficient logical qubit~\cite{Leghtas2013,Mirrahimi2014}. Autonomously protecting the logical qubit against dephasing errors requires a four-photon driven-dissipative process, which forces the harmonic oscillator to gain and lose photons in sets of four. Combining such stabilization with correction against photon loss errors using quantum nondemolition parity measurements~\cite{Lutterbach1997,Sun2014,Ofek2016,Rosenblum2018} results in complete first-order quantum error correction (QEC). 

One approach for engineering such a four-photon driven-dissipative process has been proposed in~\cite{Mundhada2017}. The idea is to implement a six-quanta process that exchanges four photons of a high-Q resonator mode $a$ (destruction operator $\bfa$) with two excitations of a transmon mode $b$ (eigen states $\ket{g}$, $\ket{e}$, $\ket{f}$) and vice versa, corresponding to an effective interaction given by $\bfa^4\ket{f}\bra{g}+\bfa^{\dagger 4}\ket{g}\bra{f}$ (see Fig.~1a). Adding a two-excitation drive and dissipation on the transmon, by employing a combination of techniques demonstrated in references~\cite{Geerlings2013,Leghtas2015}, will then result in a four-photon driven-dissipative process on the high-Q resonator. The implementation of $\bfa^4\ket{f}\bra{g}+\bfa^{\dagger 4}\ket{g}\bra{f}$ interaction requires a Raman-assisted cascading~\cite{Steck2007} of two four-wave mixing interactions, each of which exchanges two resonator photons with \textcolor{\reviewcolor}{a virtual (non-energy-conserving) excitation} in the transmon mode and a pump photon, and vice versa. \textcolor{\reviewcolor}{This transition through the virtual state plays a vital role of cascading the two nonlinear processes, and giving an effective higher-order process. On the other hand, mediating the transition through an eigen-state of the system will result in two individual processes in series, instead of a higher-order nonlinearity. Additionally, the virtual state also helps in suppressing the decoherence errors induced by the finite life-time of the transmon mode.}  

Raman transitions using linear processes~\cite[Ch.~6]{Steck2007} or a combination of one linear and one nonlinear process~\cite{Vool2018} have been previously demonstrated. Our implementation of the $\bfa^4\ket{f}\bra{g}+\bfa^{\dagger 4}\ket{g}\bra{f}$ interaction, however, requires the cascading of two nonlinear multi-quanta processes. In our experiment we show that not only the Raman-assisted cascading of nonlinear processes is feasible, but also the magnitude of the effective interaction can be made much larger than the damping rates of the high-Q modes, hence, generating a useful interaction for QEC. In principle, the same driven-dissipative process could instead be realized by using a six-wave mixing term in the Josephson cosine potential, addressed using an off-resonant pump. However, the currently achievable magnitude of the six-wave mixing term, obtained from expanding the Josephson cosine potential, is small compared to the dissipation rates of the system and other spurious terms present in the Hamiltonian~\cite[Sec.~I\,C]{SI}. Hence, Raman-assisted virtual cascading of low-order mixing processes is essential for enhancing the strength of the desired four-photon driven-dissipative process for hardware efficient QEC.

\textcolor{\reviewcolor}{This paper is organized as follows: Section~\ref{sec:exact_process} is dedicated to experimental demonstration of the cascaded higher-order process. Specifically, subsections \ref{sec:sys_details} and \ref{sec:tuneup} describe the experimental setup and the initial tuneup of the cascaded process, while, subsections \ref{sec:tomo} and \ref{sec:wigners} discuss the tomography of the cascaded process. In section~\ref{sec:discussion} we discuss some limitations of our current experiment and give future directions, followed by conclusions in section~\ref{sec:conclusion}.}

\section{Experimental demonstration}
\label{sec:exact_process}
In order to demonstrate the feasibility of cascading nonlinear processes through virtual states, our experiment focuses on the Raman-assisted $\ket{g4}\leftrightarrow\ket{f0}$ transition as explained in Fig.~1b (see figure caption for explanation). This transition is a precursor to the aforementioned $\bfa^4\ket{f}\bra{g}+\bfa^{\dagger 4}\ket{g}\bra{f}$ process which requires the $\ket{g,n}\leftrightarrow\ket{f,n-4}$ transitions to all occur simultaneously. As shown in Fig.~1b, the system is initialized in the $\ket{f0}$ state. The two pumped processes, one connecting $\ket{f0}$ to a virtual state close to $\ket{e2}$ with the rate $g_1$ and the other one connecting the virtual state to $\ket{g4}$ with the rate $g_2$, are nonlinear four-wave mixing processes. The frequencies of the two pumps involved (see Fig.~1c) are
\begin{align}
\omega_{p1}&=2\tilde\omega_a -\tilde\omega_b + \chi_{bb} - 2\chi_{ab}+\Delta\, \nonumber\\
\omega_{p2}&=2\tilde\omega_a -\tilde\omega_b + 2\chi_{ab}-\Delta\,, \label{eq:pump_frequencies}
\end{align}
where $\tilde\omega_{a/b}$ are the Stark shifted frequencies of the high-Q resonator and the transmon mode in presence of the pumps, $\chi_{ab}$ is the cross-Kerr and $\chi_{bb}$ is the self-Kerr of the transmon mode. The effective Hamiltonian of the system to second-order in the rotating wave approximation (RWA)~\cite{Mirrahimi2015} is
\begin{align}
\frac{H_{\rm eff}}{\hbar} \cong g_{\rm 4ph} \left(\ket{g4}\bra{f0}+\ket{f0}\bra{g4}\right),
\end{align}
where $g_{\rm 4ph}$ is the magnitude of the cascaded process, given by
\begin{align}
g_{\rm 4ph} = \sqrt{48}g_1 g_2\left(\frac{1}{\Delta}-\frac{1}{\chi_{bb}-4\chi_{ab}+\Delta}\right). \label{eq:g4ph}
\end{align}
For the effective Hamiltonian to be valid, one has to choose the parameters such that $|g_{1,2}|\ll \Delta$, since, as is ubiquitous in Raman transitions, the leakage rate to the intermediate state ($\ket{e2}$ in our case) is directly proportional to the ratios $|\frac{g_{1,2}}{\Delta}|^2$. Detailed derivation and discussion of the effective Hamiltonian is given in~\cite[Sec.~I\,B]{SI}.

\begin{figure}[t!]
\centering
\includegraphics[width=0.45\textwidth]{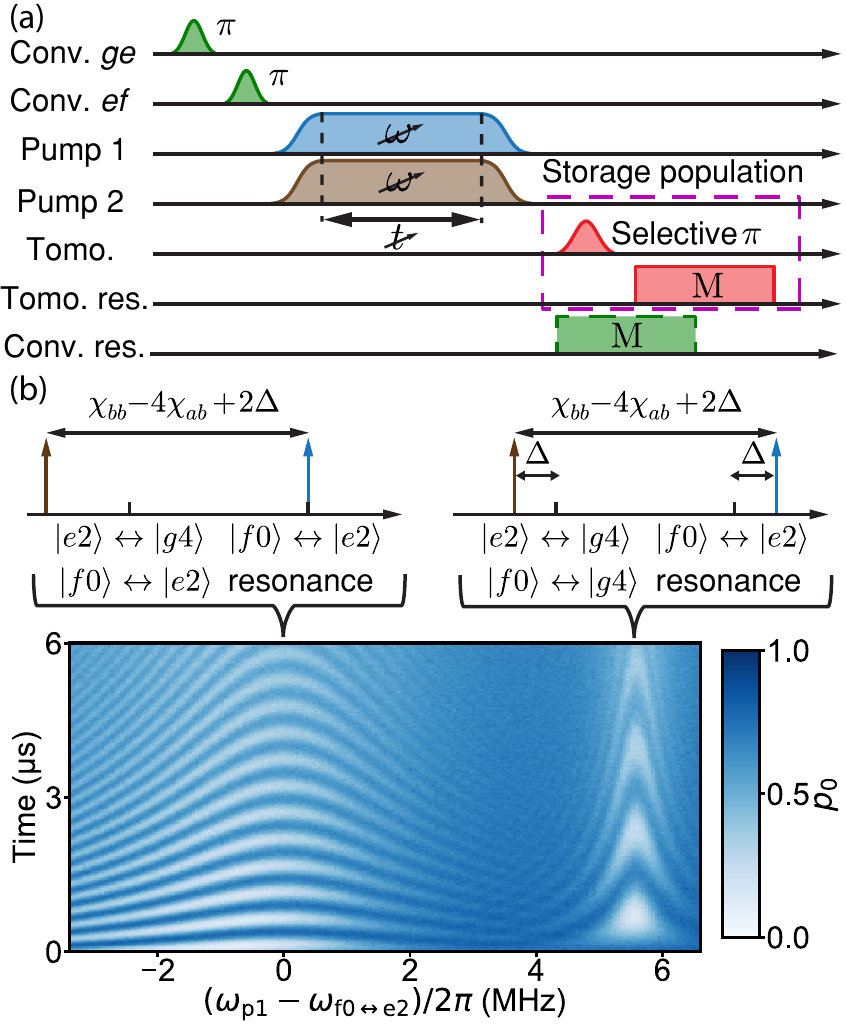}
\caption{\textbf{Pulse sequence and Rabi oscillations of the cascading process.} (a) Pulse sequence used for locating the $\ket{f0}\leftrightarrow\ket{g4}$ resonance of the system. The system is initialized in $\ket{f0}$ by using $\pi$-pulses on $\ket{g}\leftrightarrow\ket{e}$ and $\ket{e}\leftrightarrow\ket{f}$ transitions. Following this, the two pumps are applied with varying frequency and duration. The frequency difference of the two pumps is maintained constant at $\chi_{bb}-4\chi_{ab}+2\Delta$. Finally an indirect measurement of the storage resonator population is performed using a photon-number selective $\pi$-pulse on the tomography transmon and a measurement pulse on the tomography resonator. Optionally, a measurement of the conversion transmon state can also be performed using a measurement pulse on the conversion resonator. (b) Rabi oscillations in the population of Fock state $\ket{0}$ ($p_0$, colorbar). The x-axis shows the detuning of pump 1 from the $\ket{f0}\leftrightarrow\ket{e2}$ transition, the y-axis shows the duration for which the two pumps are applied. The frequency landscape above the data explains the origin of the two chevron like features.
\label{fig:figure_2}}
\end{figure}

\subsection{System details}
\label{sec:sys_details}
The experimental setup for testing our transition requires (i) a high-Q resonator, (ii) a transmon mode for the conversion process, and (iii) a second transmon mode to perform Wigner tomography~\cite{Vlastakis2013} of the resonator. In addition, we need to be able to couple pumps strongly with the conversion transmon, while maintaining the quality factor of various modes of the system. The high-Q storage resonator ($T_1=76\,\mathrm{\mu s}$) is realized as a high purity aluminum, $\lambda/4$-type, post-cavity~\cite{Reagor2013} with frequency $\omega_a/2\pi= \SI{8.03}{\giga\hertz}$ (see Fig.~1c). The resonator is dispersively coupled to two transmons as shown in Fig.~1c. The transmon in the conversion arm has a resonance frequency $\omega_b/2\pi = \SI{5.78}{\giga\hertz}$, anharmonicity $\chi_{bb}/2\pi=\SI{122.6}{\mega\hertz}$ and a cross-Kerr of $\chi_{ab}/2\pi=\SI{7.4}{\mega\hertz}$ with the high-Q resonator. The $T_1$ and $T_2$ of the conversion transmon are $50\,\mathrm{\mu s}$ and $7.6 \,\mathrm{\mu s}$ respectively. The second transmon is employed to perform Wigner tomography on the storage resonator and has a cross-Kerr of $\SI{1.1}{\mega\hertz}$ with it. Both transmons are coupled to low-Q resonators through which we perform single-shot measurements of the transmon state (see~\cite[Sec.~II\,A]{SI} for remaining system parameters). In the case of the conversion transmon, the measurement distinguishes, in single-shot, between the first three states $\ket{g}$, $\ket{e}$ and $\ket{f}$. The enclosure of the high-Q resonator acts as a rectangular waveguide high-pass filter with a cutoff at $\sim\SI{9.5}{\giga\hertz}$. Since the two pump frequencies, $\omega_{p1}/2\pi=\SI{10.397}{\giga\hertz}$ and $\omega_{p2}/2\pi=\SI{10.294}{\giga\hertz}$, are above the cutoff, they are applied through the strongly coupled (waveguide mode $Q\le 100$) pin at the top. The high-Q resonator and the transmon modes are below the cutoff and are thus protected from relaxation through this pin. 

\begin{figure}[t!]
\centering
\includegraphics[width=0.45\textwidth]{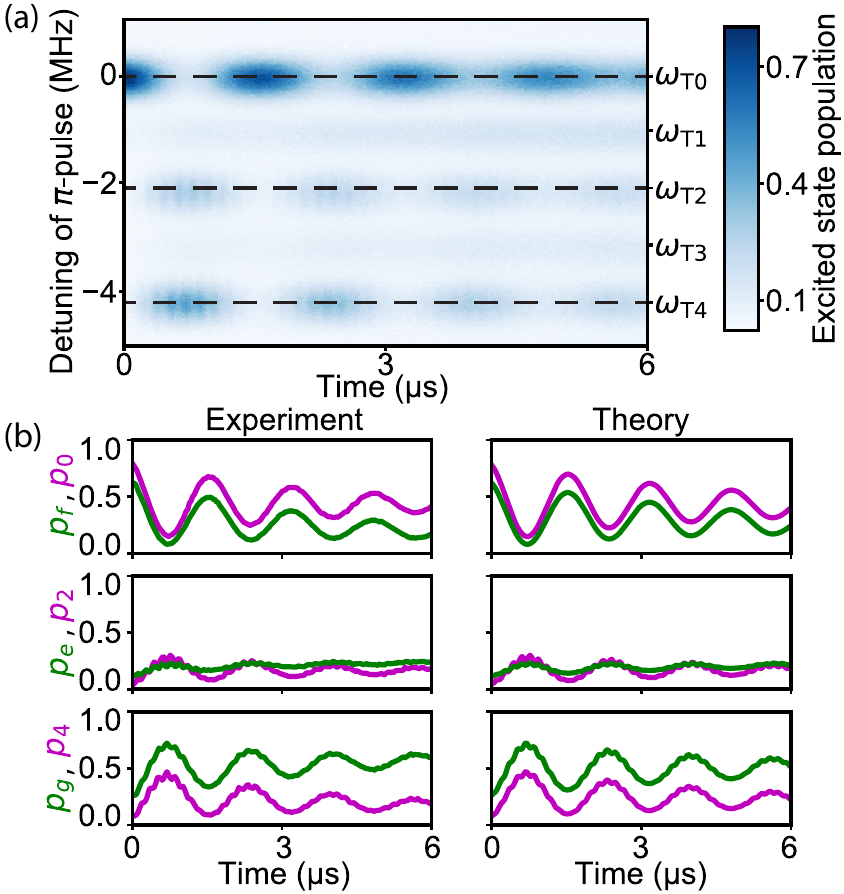}
\caption{\textbf{Partial tomography of $\ket{f0}\leftrightarrow\ket{g4}$ oscillations as a function of time.} The system is prepared in $\ket{f0}$ and the two pumps are applied for a variable period of time on resonance with the $\ket{f0}\leftrightarrow\ket{g4}$ transition. Following this, a selective pulse with a variable frequency is applied on the tomography transmon enabling an indirect measurement of various Fock state populations of the storage resonator. (a) Excited state population of tomography transmon (colorbar) versus pump duration (x-axis) and the detuning of the selective $\pi$-pulse on the tomography transmon (y-axis). The y-axis on the right shows the frequency of the tomography transmon ($\omega_{Tn}$) conditioned on the number of photons $n$ in the storage mode.\iffalse The oscillations between the states $\ket{0}$ and $\ket{4}$ are clear from the $\omega_{T0}$ and $\omega_{T4}$ lines with some leakage to the state $\ket{2}$ ($\omega_{T2}$ line) due to finite detuning $\Delta$. A small population is seen in the $\ket{1}$ and $\ket{3}$ lines due to the finite coherence time of the storage resonator. \fi (b) From top to bottom, $\ket{0}$, $\ket{2}$ and $\ket{4}$ Fock state populations (magenta), measured along the dashed lines shown in panel (a). Independently measured populations in $\ket{f}$, $\ket{e}$ and $\ket{g}$ states of the conversion mode (green) are also plotted, respectively, from top to bottom. The plots on the left are experimental data and the ones on the right are obtained from numerical simulation~\cite[Sec.~IV]{SI}. \label{fig:figure_3}}
\end{figure}

\subsection{Spectroscopic tuneup}
\label{sec:tuneup}
In order to locate the correct pump frequencies for the transition of interest, we use the pulse sequence shown in Fig.~2a. The system is initialized in $\ket{f0}$ and the two pumps are applied for a variable period of time. The pump frequencies are swept such that the frequency difference is maintained constant at $\omega_{p1}-\omega_{p2}=\chi_{aa}-4\chi_{ab}+2\Delta$. We choose $\Delta/2\pi=\SI{5.1}{\mega\hertz}$ and $g_{1,2}/2\pi\sim \SI{0.5}{\mega\hertz}$. The rising and falling edges of the pump pulses are smoothed using a hyperbolic tangent function with a smoothing time of $\SI{192}{\nano\second}$. These parameters are empirically optimized to reduce the leakage to the $\ket{e2}$ state while achieving a $g_{\rm 4ph}$ that is an order of magnitude faster than the decoherence rates of the system. The resulting resonator state is characterized by applying a photon-number selective $\pi$-pulse~\cite{Leghtas2013a} on the tomography transmon. The pulse has a gaussian envelope of width $\sigma_{\rm sel}=\SI{480}{\nano\second}$ (total length $4\sigma_{\rm sel}$), resulting in a pulse bandwidth of $\sim\SI{332}{\kilo\hertz}$, which is less than the cross-Kerr between the tomography transmon and the high-Q resonator. As a result the tomography transmon is excited only when the storage resonator is in $\ket{0}$. Finally, the state of the tomography transmon is measured. An optional single-shot measurement of the conversion transmon can also be performed as indicated by the dashed green measurement pulse in Fig.~2a.

\begin{figure}[t!]
\centering
\includegraphics[width=0.45\textwidth]{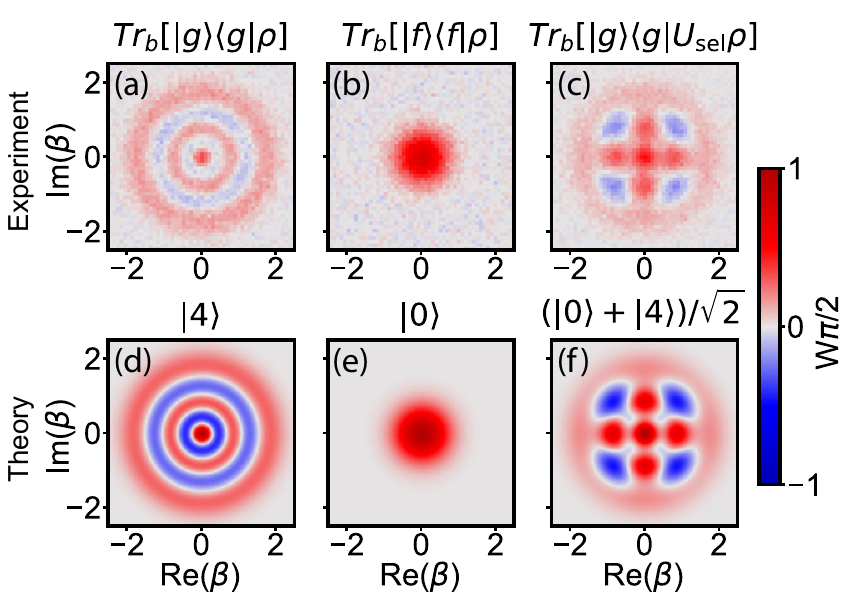}
\caption{\textbf{Conditional Wigner tomography of the storage resonator after a quarter period of the $\ket{f0}\leftrightarrow\ket{g4}$ oscillation.} After quarter period of $\ket{f0}\leftrightarrow\ket{g4}$ oscillation the system is in the state $\left(\ket{f0}+\ket{g4}\right)/\sqrt{2}$. (a, b) show experimental Wigner function of the storage resonator after post-selecting the conversion mode in the $\ket{g}$ and $\ket{e}$ states. This leaves the storage resonator in Fock states $\ket{4}$, $\ket{0}$ respectively. (d, e) show the ideal Wigner functions of Fock states $\ket{4}$, $\ket{0}$ for comparison. (c) Wigner function of the resonator after photon-number selective $\pi$-pulses from $\ket{f0}$ to $\ket{e0}$ and $\ket{e0}$ to $\ket{g0}$ (indicated by $U_{\rm sel}$) and post-selecting the conversion transmon in $\ket{g}$. Comparing (c) with the ideal Wigner function of $(\ket{0}+\ket{4})/\sqrt{2}$ state in (f), shows that the storage resonator is in a coherent superposition of $\ket{0}$ and $\ket{4}$, thus indicating that the $\ket{f0}\leftrightarrow\ket{g4}$ oscillations are coherent.
\label{fig:figure_4}}
\end{figure}

The outcome of the described measurement is shown in Fig.~2b. The population fraction of the Fock state $\ket{0}$ is plotted as a function of the duration for which the pump pulses are applied and the detuning of the first pump $\omega_{p1}$ from the $\ket{f0}\leftrightarrow\ket{e2}$ transition. The data displays Rabi oscillations arising from two processes. The one on the left occurs when pump 1 is resonant with the $\ket{f0}\leftrightarrow\ket{e2}$ transition. The one on the right corresponds to the two pumps being equally detuned from the $\ket{f0}\leftrightarrow\ket{g2}$ and $\ket{e2}\leftrightarrow\ket{g4}$ transitions. This is the Raman-assisted $\ket{f0}\leftrightarrow\ket{g4}$ transition of interest. The resulting chevron pattern for this transition is narrower since the cascaded transition occurs at a slower rate than the $\ket{f0}\leftrightarrow\ket{g2}$ transition. From the frequency of the oscillations we extract $g_{\rm 4ph}/2\pi=\SI{0.32}{\mega\hertz}$. In separate experiments, we accurately characterize the pump strengths $g_1/2\pi=\SI{0.53}{\mega\hertz}$ and $g_2/2\pi=\SI{0.48}{\mega\hertz}$ by measuring the Stark shifts of the conversion transmon, when the pumps are applied separately at their respective resonance conditions for the $\ket{f0}\leftrightarrow\ket{g4}$ transition~\cite[Sec.~III\,B]{SI}. This eliminates any frequency dependent attenuation of pump strengths due to the dispersion in the input lines. For these parameters, Eq.~\eqref{eq:g4ph} predicts a $g_{\rm 4ph}/2\pi$ of $\SI{0.33}{\mega\hertz}$, in close agreement with the measured value. 

\subsection{Partial tomography of $\ket{f0}\leftrightarrow\ket{g4}$ process}
\label{sec:tomo}
Having found the desired $\ket{f0}\leftrightarrow\ket{g4}$ process, we fix our pump frequencies to be resonant with this transition and proceed to characterize the populations of different Fock states of the storage resonator. These are obtained by varying the frequency at which the photon-number selective pulse on the tomography transmon is applied. The result of this measurement is plotted in Fig.~3a. The population fractions of various Fock states are inferred by taking cross-sections at the resonance frequency of the tomography transmon conditioned on the number of photons in the high-Q resonator. The resonator oscillates between $\ket{0}$ and $\ket{4}$ with some leakage to $\ket{2}$ due to the finite detuning $\Delta$ from $\ket{e2}$ (see the $\omega_{T0/2/4}$ lines in Fig.~3a). The population appearing in $\ket{1}$ and $\ket{3}$ is due to finite energy relaxation time of the resonator mode. The evolution of the $\ket{0}$, $\ket{2}$ and $\ket{4}$ state populations of the storage resonator and the $\ket{f}$, $\ket{e}$, $\ket{g}$ state populations of the conversion transmon as a function of time are plotted in the first column of Fig.~3b. The conversion transmon populations are measured independently using the dashed-green measurement pulse shown in Fig.~2a. The respective populations oscillate in phase with each other as expected. The amplitude of the oscillations is limited by the $T_2$ of the conversion qubit and the contrast of the two measurements. We are also able to resolve an envelope of fast oscillations in the populations of $\ket{e}$, $\ket{g}$ and $\ket{2}$, $\ket{4}$ states. These are expected for a Raman transition and occur at a rate given by the detuning $\Delta$. The plots in the second column of Fig.~3b show numerical data obtained from simulating Lindblad master equation of the system~\cite[Sec.~IV]{SI}. The contrast of the simulation is scaled by the measurement contrast of the experimental system. The simulation reproduces the experimental results well, including the fast oscillations found in the data.

\subsection{Coherence of $\ket{f0}\leftrightarrow\ket{g4}$ process}
\label{sec:wigners}
Finally, in order to demonstrate that the oscillations are coherent, we stop the oscillations after a quarter of a period ($\SI{372}{\nano\second}$). This is expected to prepare a coherent superposition of $\ket{f0}$, $\ket{g4}$ given by $\left(\ket{f0}+\ket{g4}\right)/\sqrt{2}$. We experimentally characterize the state of the system by performing Wigner tomography of the resonator, conditioned on conversion transmon states. As expected, the resonator ends up in Fock state $\ket{4}$ ($\ket{0}$) when the conversion transmon is post-selected in $\ket{g}$ ($\ket{f}$) as shown by Fig.~4a (4b). Moreover, applying a photon number selective $f\rightarrow g$ pulse on the conversion transmon, conditioned on zero photons in the storage resonator, disentangles the transmon from the resonator, leaving the system in $\ket{g}\otimes\left(\ket{0}+\ket{4}\right)/\sqrt{2}$. The Wigner function of the resonator after post-selecting the conversion transmon in $\ket{g}$, shown in Fig.~4c, depicts a $\left(\ket{0}+\ket{4}\right)/\sqrt{2}$ state, thus proving that the oscillations are coherent. For comparison, the ideal Wigner functions of $\ket{4}$, $\ket{0}$ and $\left(\ket{0}+\ket{4}\right)/\sqrt{2}$ are shown in panels d, e and f of Fig.~4 respectively. It is also interesting to note that $\left(\ket{0}+\ket{4}\right)/\sqrt{2}$ is one of the logical states of binomial QEC codes~\cite{Michael2016}.

\section{Discussion}
\label{sec:discussion}
\textcolor{\reviewcolor}{While we have demonstrated a six-quanta $\ket{g4}\leftrightarrow\bra{f0}$ transition, autonomous QEC requires a $\bfa^4\ket{f}\bra{g}+\bfa^{\dagger 4} \ket{g}\bra{f}$ process, where all of the $\ket{gn}\leftrightarrow\ket{f(n-4)}$ transitions are resonant simultaneously. This can be accomplished by making the strength of the pumped processes $g_{1,2}$, higher than the cross-Kerr terms $\chi_{ab}$ between the storage resonator and the conversion transmon. However, such pump strengths are not achievable in our current system, due to spurious transitions induced by strong pump strengths, similar to those seen in references~\cite{Sank2016,Lescanne2019}. This limitation, however, should not discourage future applications, since, there have been proposals to increase tolerance for the pump strengths by shunting the transmon with a linear inductor~\cite{Verney2019} or using flux-biased circuits to cancel cross-Kerr between modes~\cite{Elliott2018}.} 

\textcolor{\reviewcolor}{The leakage to the intermediate state $\ket{e(n-2)}$ could be another limitation for QEC applications. In future iterations of our experiment, this leakage can be minimized by increasing the detuning and making the pulses more adiabatic, albeit at the cost of making the overall process slower. It is also possible to use pulse shaping techniques like stimulated Raman adiabatic passage (STIRAP)~\cite[Ch. 6.2.3]{Steck2007} to implement this transition without any leakage. The effect of this leakage on the error-correction protocol is discussed at length in Ref.~\cite{Mundhada2017}. Moreover Ref.~\cite{Albert2019} details an alternative QEC scheme which uses a similar driven-dissipative process, however, it is insensitive to leakage to the $\ket{e,n-2}$ state.}\\

\section{Conclusion}
\label{sec:conclusion}
In conclusion, we have shown that nonlinear processes can be cascaded through a virtual state to engineer higher-order nonlinear Hamiltonians. \textcolor{\reviewcolor}{The rate of this highly nonlinear transition is faster than the decoherence rates. The oscillations are coherent and follow the theoretical predictions closely. The demonstrated $\ket{g4}\leftrightarrow\ket{f0}$ oscillations are a precursor to the implementation of the complete $\bfa^4\ket{f}\bra{g}+\bfa^{\dagger 4}\ket{g}\bra{f}$ Hamiltonian, which is an important component of hardware efficient quantum error correction using Schr\"odinger cat-states.}

Moreover, while three- and four-wave mixing processes have played a key role in cQED applications~\cite{Vijay2009,Abdo2011,Macklin2015,White2015,Narla2016,Frattini2017,Metelmann2017}, \textcolor{\reviewcolor}{many proposals will benefit from increasingly higher-order nonlinear interactions~\cite{Mamaev2018,Kapit2016,Lihm2018,Albert2019}}. We have accomplished a deeper goal of verifying that higher-order nonlinear interactions can indeed be engineered by cascading lower-order nonlinear processes. As shown in~\cite[Sec.~I\,A]{SI}, it is possible to cascade any two processes through a virtual state, as long as the commutator of the operators that describe the processes is the operator describing the desired higher-order process. Therefore, such cascading could be useful for the broader field of quantum optics and quantum control. Additionally, the possibility of cascading indicates that advanced techniques like GRAPE (gradient-ascent pulse engineering)~\cite{Khaneja2005,Fouquieres2011} could utilize pulses addressing nonlinear processes to gain additional control knobs over the system, thus potentially increasing the speed and fidelity of the engineered unitary operations.

\section*{Acknowledgements}
The authors acknowledge support from ARO grant number W911NF-14-1-0011 and W911NF-18-1-0212. Use of fabrication facilities was supported by the Yale Institute for Nanoscience and Quantum Engineering(YINQE) and the Yale SEAS cleanroom.

\bibliography{prospectus_citations}

\end{document}


\onecolumngrid
\begin{center}
\textbf{\large Supplemental Materials for ``Experimental implementation of a Raman-assisted six-quanta process"}
\end{center}

\section{Theory}

Here, we discuss the necessary conditions for Raman-assisted virtual cascading and present the calculations pertaining to the $\ket{f0}\leftrightarrow\ket{g4}$ oscillations demonstrated in the experiment. Additionally, we compare the magnitude of this process with the estimated magnitude of currently achievable six-wave mixing processes.

\subsection{Designing a Raman-assisted higher-order process}
\label{sec:general_raman}

In this subsection we use the expressions for second-order rotating wave approximation (RWA)~\cite{Mirrahimi2015} to obtain some pointers towards designing Raman-assisted higher-order processes. Consider a Hamiltonian in an interaction picture with respect to the diagonal part, given by
\begin{align}
\frac{H_{I}(t)}{\hbar} = \frac{H_c}{\hbar}+g_1 e^{i \Delta t} \mathbf{A}_1 + g_1^* e^{-i \Delta t} \mathbf{A}_1^\dagger + g_2 e^{-i \Delta t} \mathbf{A}_2 + g_2^* e^{i \Delta t} \mathbf{A}_2^\dagger. \label{eq:general_hamiltonian}
\end{align}
Here, $H_c$ is time-independent part of $H_I(t)$ and $\mathbf{A}_1$, $\mathbf{A}_2$ are operators describing off-diagonal interactions available in the system. \textcolor{\reviewcolor}{The specific expressions for $\mathbf{A}_1$ and $\mathbf{A}_2$, in the case of our experiment, are considered in a latter section.} In the given rotating frame of $H_I(t)$, the two processes are detuned by $+\Delta$ and $-\Delta$ respectively. The effective Hamiltonian to the second-order in RWA is given by
\begin{align}
H_{\rm RWA} &= \overline{H_I(t)} - i\overline{\left(H_I(t)-\overline{H_I(t)}\right)\int\mathrm{d}t \left(H_I(t)-\overline{H_I(t)}\right)}
\end{align} 
where $\overline{H(t)}=\lim_{T\rightarrow \infty} \frac{1}{T}\int_0^T H(t)\mathrm{d}t$.  Applying this to the Hamiltonian in Eq.~\ref{eq:general_hamiltonian} we obtain
\begin{align}
\frac{H_{\rm eff}}{\hbar}= \frac{H_c}{\hbar}+\frac{|g_1|^2}{\Delta}\left[\mathbf{A}_1,\mathbf{A}_1^\dagger\right]-\frac{|g_2|^2}{\Delta}\left[\mathbf{A}_2,\mathbf{A}_2^\dagger\right]+\frac{g_1g_2}{\Delta}\left[\mathbf{A}_1,\mathbf{A}_2\right]-\frac{g_1^*g_2^*}{\Delta}\left[\mathbf{A}_1^\dagger,\mathbf{A}_2^\dagger\right]. \label{eq:effective_hamiltonian_general}
\end{align} 
This reveals effective interactions given by the commutation relations of the operators $\mathbf{A}_1$ and $\mathbf{A}_2$. Therefore, in order to design a Raman-assisted higher-order process we have to use the following three principles.
\begin{itemize}
\item Select the lower-order processes such that their commutation relation is a non-zero operator describing the required higher-order process.
\item Design the lower-order processes to be oscillating with equal and opposite frequencies $\Delta$ so that their product survives the second order RWA.
\item \textcolor{\reviewcolor}{Engineer the time independent part $H_C$ in Eq.\eqref{eq:general_hamiltonian} to cancel the effect of the unwanted resonant terms in $H_{\rm eff}$ as well as effects arising at higher-orders in perturbation theory (not shown).}
\end{itemize}
Another issue to keep in mind is the validity of second-order RWA. Eq.~\ref{eq:effective_hamiltonian_general} is a good approximation only when $\frac{|g_1|}{\Delta},\frac{|g_2|}{\Delta}\ll 1$. In general, along with the interactions given in $H_{\rm eff}$, the two individual processes described by $A_1 , A_2$ are also off-resonantly enabled, leading to leakages corresponding to $A_1, A_2$ transitions. These leakages can be minimized by selecting smaller values of $\frac{|g_{1,2}|}{\Delta}$ albeit at the cost of slowing the desired effective process as well.

\subsection{Calculations for $\ket{f0}\leftrightarrow\ket{g4}$ process}
\label{sec:g4_f0_calculations}

The calculation for a general process described by $\bfa^4 \ket{f}\bra{g}+\bfa^{\dagger 4} \ket{g}\bra{f}$ is done in \cite{Mundhada2017}. As mentioned in the main text, this process requires all $\ket{gn}\leftrightarrow\ket{fn-4}$ transitions to be resonant simultaneously. However, when the cross-Kerr between the cavity mode and the transmon mode is large compared to the strengths of the individual four-wave mixing processes ($g_{1,2}$), as is the case in our current system, the resonant frequencies of $\ket{gn}\leftrightarrow\ket{fn-4}$ transitions depends on $n$. Hence, the process that we have demonstrated is photon-number selective. Here we present the pertinent calculations where we ignore the presence of the tomography transmon. Following the analysis in \cite{Mundhada2017} the Hamiltonian of the system in a displaced frame where the drives have been absorbed in the cosine potential is
\begin{align}
\frac{H}{\hbar} = \omega_a \bfa^\dagger \bfa + \omega_b \bfb^\dagger \bfb - \frac{E_J}{\hbar} \left[\cos\left(\mathbf{\Phi}(t)\right)+\frac{\mathbf{\Phi}^2(t)}{2!}\right], \label{eq:displaced_hamiltonian}
\end{align}
\textcolor{\reviewcolor}{where $\bfb$ is the destruction operator corresponding to the conversion transmon and} $\Phi(t)$ is the phase across the Josephson junction of the conversion transmon. The expression for $\Phi(t)$ is given by
\begin{align}
\mathbf{\Phi}(t)=\phi_a \left(\bfa+\bfa^\dagger\right)+\phi_b \left(\bfb+\bfb^\dagger\right) + \phi_b \sum_{k=1,2} \xi_k \exp\left(-i\omega_{pk}t\right)+\xi_k^* \exp\left(i\omega_{pk}t\right), \nonumber
\end{align}
and $E_J$ is the Josephson energy. The ratios $\phi_{a,b}$ are the dimensionless inductive participation amplitudes of the cavity and the transmon modes in the junction with typical magnitudes of $\phi_a^2 \sim 10^{-3}$ and $\phi_b^2 \sim 10^{-1}$ respectively. The two applied pump drives are at the frequencies $\omega_{pk}$ defined as 
\begin{align}
\omega_{p1}&=2\tilde{\omega}_a-\tilde{\omega}_b+\chi_{bb}-2\chi_{ab}+\Delta+\delta\, \nonumber\\
\omega_{p2}&=2\tilde{\omega}_a-\tilde{\omega}_b+2\chi_{ab}-\Delta+\delta\,. \nonumber
\end{align}
The frequencies $\tilde{\omega}_a, \tilde{\omega}_b$ are the Stark shifted frequencies of modes $a$ and $b$ respectively. The shift $\delta$ is added to both pump frequencies in order to account for higher-order frequency shifts in the system. Additionally, for the purpose of this calculation, we have assumed that the pumps only couple to mode $b$. This assumption does not lead to any loss of generality since the coupling to mode $a$ can be effectively absorbed in the time dependent part of $\Phi(t)$ with a slight modification to $\xi_k$.  

Expanding the Hamiltonian in Eq.~\ref{eq:displaced_hamiltonian} to the fourth order in cosine expansion and keeping only the terms that will survive after second order RWA, we get
\begin{align}
\frac{H_{\rm sys}}{\hbar}=\frac{H_0}{\hbar} + \left(g_1 e^{-i\omega_{p1} t}+g_2 e^{-i\omega_{p2} t}\right) \bfa^2\bfb^\dagger + \left(g_1^* e^{i\omega_{p1} t}+g_2^* e^{i\omega_{p2} t}\right) \bfa^{2\dagger}\bfb. \label{eq:H_sys}
\end{align}
where 
\begin{align}
\frac{H_0}{\hbar}= \tilde\omega_a \bfa^\dagger \bfa + \tilde\omega_b \bfb^\dagger \bfb - \chi_{ab} \bfa^\dagger \bfa \bfb^\dagger \bfb - \frac{\chi_{aa}}{2} \bfa^{\dagger 2} \bfa^2 - \frac{\chi_{bb}}{2} \bfb^{\dagger 2} \bfb^2 \nonumber
\end{align}
and
\begin{align}
g_{1,2} = -\frac{E_J \phi_a^2\phi_b^2}{2}\xi_{1,2} = -\frac{\chi_{ab}}{2}\xi_{1,2}. \label{eq:g1_g2_magnitude}
\end{align}
The expressions for the Stark shifted frequencies of the modes are
\begin{align}
\tilde{\omega}_a &=\omega_a - \left(|\xi_1|^2+|\xi_2|^2\right)\chi_{ab}\, \nonumber \\
\tilde{\omega}_b &=\omega_b - 2 \left(|\xi_1|^2+|\xi_2|^2\right)\chi_{bb}\,\label{eq:Stark_shifted_freqs}
\end{align}
where $\omega_{a,b}$ are the bare frequencies. Going into the rotating frame with respect to $H_0/\hbar + \frac{\chi_{aa}}{2} \bfa^{\dagger 2}\bfa^2 - \delta \bfb^\dagger \bfb$ we get the Hamiltonian in the interaction picture
\begin{align}
\frac{H_{\rm I}}{\hbar} = &-6\chi_{aa}\ket{g4}\bra{g4}-\left(\chi_{aa}+\delta\right)\ket{e2}\bra{e2}-2\delta\ket{f0}\bra{f0}\nonumber\\
& +\sqrt{4} \left[g_1 \exp\left(i\Delta t\right) + g_2 \exp\left(-i(\chi_{bb}-4\chi_{ab}+\Delta)t\right)\right] \ket{f0}\bra{e2} + \mbox{h.c.} \nonumber \\
&+ \sqrt{12} \left[g_1  \exp\left(i(\chi_{bb}-4\chi_{ab}+\Delta)t\right) + g_2 \exp\left(-i\Delta t\right)\right] \ket{e2}\bra{g4} + \mbox{h.c.}\,,
\end{align}
where $\mbox{h.c.}$ indicates Hermitian conjugate. Comparing this expression with the expression given in Eq.~\eqref{eq:general_hamiltonian}, we can infer that the first row is the time independent part $H_c$ and, $\ket{f0}\bra{e2}$, $\ket{e2}\bra{g4}$ are the individual $A_1$, $A_2$ processes in Eq.~\eqref{eq:general_hamiltonian}. The other terms in the Hamiltonian do not contribute after second order RWA and hence are ignored. Finally, performing the RWA as specified in Sec.~\ref{sec:general_raman}, we get the effective Hamiltonian
\begin{align}
\frac{H_{\rm eff}}{\hbar} = g_{\rm 4ph} & \left(\ket{g4}\bra{f0}+\ket{f0}\bra{g4}\right) \nonumber \\
&+\left(\frac{12|g_2|^2}{\Delta}-\frac{12|g_1|^2}{\chi_{bb}-4\chi_{ab}+\Delta}-6\chi_{aa}\right)\ket{g4}\bra{g4} \nonumber\\
&-\left(\frac{12|g_2|^2+4|g_1|^2}{\Delta}-\frac{4|g_2|^2+12|g_1|^2}{\chi_{bb}-4\chi_{ab}+\Delta}+\chi_{aa}+\delta\right)\ket{e2}\bra{e2}\nonumber\\
&+\left(\frac{4|g_1|^2}{\Delta}-\frac{4|g_2|^2}{\chi_{bb}-4\chi_{ab}+\Delta}-2\delta\right)\ket{f0}\bra{f0}\,.
\end{align}
The first row in the above equation is the desired effective interaction. The magnitude of $g_{\rm 4ph}$ has been quoted in Eq.~(3) of the main text. The other terms in the Hamiltonian are higher-order frequency shifts introduced by the pumps. We compensate for these shifts in the experiment by sweeping the pump frequencies while keeping the difference between them constant at $\chi_{bb}-4\chi_{ab}+2\Delta$. This common shift of pump frequencies given by $\delta$, amounts to
\begin{align}
\delta = 3\chi_{aa} + \left(\frac{2|g_1|^2-6|g_2|^2}{\Delta}+\frac{6|g_1|^2-2|g_2|^2}{\chi_{bb}-4\chi_{ab}+\Delta}\right). \label{eq:delta}
\end{align}
It can be seen that for this value of delta, the higher-order frequency shifts introduced in $\ket{f0}$ and $\ket{g4}$ states are equal, thus making the $\ket{f0}\leftrightarrow\ket{g4}$ transition resonant.

\subsection{Comparison with the magnitude of six-wave mixing process}

As mentioned in the main text, the four-photon driven-dissipative process required to stabilize a manifold of four-component Schr\'odinger cat states can, in principle, be implemented in two distinct ways. The first way is through Raman-assisted cascading, which is the topic of exploration for our letter. The other way is using the six-wave mixing capabilities of a Josephson junction. The idea is to exchange four photons of a storage resonator with a single excitation of a Josephson junction mode such as transmon, SQUID, SNAIL~\cite{Frattini2017} etc., accompanied by a release of pump-photon, and vice versa; a five-quanta process. In this section we compare the estimated magnitude of this six-wave mixing process with that of the Raman-assisted cascading. 

The magnitude of the six-wave mixing process can be estimated by expanding the cosine potential in Eq.~\eqref{eq:displaced_hamiltonian} to the sixth-order. The expression for the rate of this interaction is
\begin{align}
g_{\rm 6-wave}=\frac{E_J}{24\hbar}\phi_a^4 \phi_b^2 \xi_{0}=\frac{\phi_a^2}{24}\chi_{ab}\xi,\label{eq:6_wave_rate}
\end{align} 
where $\xi_{0}$ is the strength of the pump addressing the five-quanta process. On the other hand, using the expressions in~\cite{Mundhada2017}, one gets the rate of the Raman-assisted $\bfa^4\bfb^{\dagger 2} + \bfa^{\dagger 4}\bfb^2$ process as 
\begin{align}
g_{\rm Raman}=\frac{\chi_{ab}\xi}{20} \left(1 - \frac{5\chi_{ab}\xi}{\chi_{bb}-4\chi_{ab}+5\chi_{ab}\xi}\right).\label{eq:raman_rate}
\end{align}
Here we have substituted $g_1=g_2=\chi_{ab}\xi/2$ and $\Delta = 10g_{1,2}=5\chi_{ab}\xi$. This maintains $\Delta\gg g_{1,2}$ for the RWA to be valid.  In order to estimate the relative strength of the processes, we use $\xi_0=\xi_1=\xi_{2}\cong 0.2$ and $\phi_a^2\cong 0.002$ obtained by using the parameters of our system as a guide. These are representative of the typical parameters achievable in resonators coupled to transmon modes. The pump strengths are chosen based on an empirically observed limit, imposed by the chaotic behavior of high-Q transmon-resonator system, at high pump powers~\cite{Sank2016,Lescanne2019}. Using Eq.~\eqref{eq:6_wave_rate} and Eq.~\eqref{eq:raman_rate}, we estimate that the Raman-assisted process will be stronger than the six-wave mixing process by about two-orders of magnitude. Moreover, assuming the validity of these rate expressions at higher pump-powers, the Raman transition dominates the six-wave mixing process till the pump strength $\xi\geq 600$. In reality the expressions shown here break down at such high pump strengths~\cite{Verney2019} and, in high-Q devices, these regimes have not been experimentally achieved yet.

\section{Experiment} 

Further details on the system parameters, experimental setup and measurement protocols involved are presented in this section.

\subsection{System parameters}

Here we give the detailed parameters of our experimental system. As mentioned in the main text, we have a high-Q storage resonator coupled to two transmon modes which are the conversion transmon and the tomography transmon. Each transmon is in turn coupled to one low-Q resonators which facilitates single-shot heterodyne measurement of the transmon. Hence, in total, we have five crucial modes in our system. Table \ref{tab:sys_params} specifies their frequencies, coherence times and coupling strengths with each other (off-diagonal $\chi$s in the table). The self-Kerr and $T_2$ are measured and specified only for the high-Q modes which are the storage resonator and the two transmon modes. In particular it is noteworthy that the modes for which the cross-Kerr are listed as NA, are indeed isolated from each other physically and hence, have negligible cross-Kerr between them. 

\begin{table}[h]
\begin{tabular} { | m{0.15\textwidth} | m{0.15\textwidth} | m{0.15\textwidth} | m{0.15\textwidth} | m{0.15\textwidth} | m{0.15\textwidth} | }
\hline
   & Storage resonator & Conversion transmon & Conversion resonator & Tomography transmon & Tomography resonator \\
\hline
Storage resonator & \begin{tabular}{l} f: $\SI{8.03}{\giga\hertz}$\\ $T_1$: $\SI{72}{\micro\second}$ \\ $T_2$: $\SI{56}{\micro\second}$ \\ $\chi_{\rm self}$: $\SI{122}{\kilo\hertz}$ \end{tabular} & $\chi$: $\SI{7.4}{\mega\hertz}$ & NA & $\chi$: $\SI{1.1}{\mega\hertz}$  & NA \\ 
\hline
Conversion transmon & $\chi$: $\SI{7.4}{\mega\hertz}$ & \begin{tabular}{l} f: $\SI{5.78}{\giga\hertz}$\\ $T_1$: $\SI{50}{\micro\second}$ \\ $T_2$: $\SI{7.6}{\micro\second}$ \\ $\chi_{\rm self}$: $\SI{122}{\mega\hertz}$ \end{tabular}  & $\chi$: $\SI{5.7}{\mega\hertz}$  & NA & NA  \\ 
\hline
Conversion resonator & NA & $\chi$: $\SI{5.7}{\mega\hertz}$  & \begin{tabular}{l} f: $\SI{9.93}{\giga\hertz}$\\ $\kappa$: $\SI{5.32}{\mega\hertz}$ \end{tabular} & NA & NA  \\ 
\hline
Tomography transmon & $\chi$: $\SI{1.1}{\mega\hertz}$  & NA & NA & \begin{tabular}{l} f: $\SI{6.36}{\giga\hertz}$\\ $T_1$: $\SI{38}{\micro\second}$ \\ $T_2$: $\SI{8.8}{\micro\second}$ \\ $\chi_{\rm self}$: $\SI{264}{\mega\hertz}$ \end{tabular} & $\chi$: $\SI{0.9}{\mega\hertz}$  \\
\hline
Tomography resonator & NA & NA & NA & $\chi$: $\SI{0.9}{\mega\hertz}$  & \begin{tabular}{l} f: $\SI{7.53}{\giga\hertz}$\\ $\kappa$: $\SI{0.38}{\mega\hertz}$ \end{tabular}  \\
\hline
\end{tabular}
\caption{System parameters}\label{tab:sys_params}
\end{table}

\subsection{Measurement setup}

The principles of our measurement setup are similar to those shown in \cite{Touzard2018}. A detailed wiring diagram has been shown in Fig. S1. The upper half contains the room temperature wiring (above $\SI{300}{\kelvin}$ dashed line) of the experiment and the lower half shows the wiring inside the dilution refrigerator. As mentioned in the main text, we have two transmon qubits and the ability to perform single-shot measurement on both qubits. The low-Q resonator coupled to the conversion transmon has a frequency of $\SI{9.93}{\giga\hertz}$ as mentioned in table \ref{tab:sys_params}. This is above the cutoff frequency of the waveguide enclosure and hence, this mode couples to the transmission line through the strongly coupled pin situated at the top of the waveguide. This coupling pin serves the dual purpose of measurement pin for the conversion transmon as well as the pin through which the off-resonant pumps are applied. Moreover, the coupling pin only addresses the waveguide mode with the polarization along the length of the pin. Hence, the applied pumps only couple to the conversion transmon while leaving the tomography transmon unperturbed. All the tones applied on this pin are combined using a directional coupler and routed to the coupling pin using a circulator. The directional coupler also sends most of the pump signal back to room temperature, hence effectively attenuating the pump tones without heating up the base plate of the dilution refrigerator. The circulator directs the reflected signal from the waveguide pin towards a Josephson parametric converter (JPC) which amplifies the signal at the conversion resonator frequency and sends it back to room temperature via circulator, isolators and a high electron mobility transistor (HEMT) amplifier placed at 4K. The coupling pin situated close to the conversion arm is weakly coupled to the system and is used to drive the conversion transmon. The pin situated on the tomography arm, however, is strongly coupled to the tomography resonator and is used for three purposes. Firstly, it is used to readout the tomography resonator in reflection. The signal is routed using two circulators to a SNAIL parametric amplifier (SPA) \cite{Frattini2018} and the amplified signal is routed through the circulator towards the output chain. The other two purposes of the tomography arm coupling pin are to address the tomography qubit as well as the storage resonator. In fact, the relaxation time of the storage resonator is limited because of the coupling to the environment via this pin. At room temperature, we have five generators to address the system and two more for powering the amplifiers. The generators addressing the conversion resonator and the storage resonator are also combined to produce a tone close to the frequencies of the pumps thus phase locking the two modes with the pumps. The other three generators are used to address the pumping resonator, the tomography qubit and the tomography resonator.

\section{Methods}

\subsection{Sample fabrication}

All the modes of the system are simulated using ANSYS HFSS and the Hamiltonian of the system is inferred using energy participation ratio black-box quantization technique~\cite{Nigg2012}. The cavity enclosure is machined into a single block of high purity aluminum in order to make a seamless re-entrant cavity \cite{Reagor2013}. The transmons are fabricated as Al/Al$\rm O_{\rm x}$/Al Josephson junctions on a c-plane double-side polished sapphire wafer using bridge-free electron beam lithography \cite{Lecocq2011}. The low-Q resonators are realized as stripline $\lambda/2$ resonators defined lithographically. The coupling pins shown in the Fig.~1d and Fig.~S1 are coaxial couplers whose coupling strength is tuned by adjusting the length of their exposed pin.

\subsection{Pump strength calibration}

Accurate measurement of the rates $g_{1}$ and $g_{2}$ is necessary for comparing our experimental data for the Raman-assisted $\ket{g4}\leftrightarrow\ket{f0}$ oscillations with the theoretical predictions (Sec.~\ref{sec:g4_f0_calculations}) and the simulation results (Sec.~\ref{sec:sim_details}). The rates $g_{1,2}$ are related to the pump strengths ($\xi_{1,2}$) as shown by Eq.~\eqref{eq:g1_g2_magnitude}. However, the pumps experience frequency dependent coupling strengths and attenuation of the input lines, changing $\xi_{1,2}$ as a function of pump frequency. In order to eliminate this frequency dependence, we calibrate the pump strengths by measuring the Stark shift of the transmon mode caused by each individual pumps (while the other pump is off), when the pump frequencies are same as those used for addressing the $\ket{g4}\leftrightarrow\ket{f0}$ transition. Using Eq.~\eqref{eq:Stark_shifted_freqs} we can relate the measured Stark shift to the pump strengths $\xi_{1,2}$ and by extension calibrate $g_{1,2}$ using Eq.~\eqref{eq:g1_g2_magnitude}. The Stark shift due to pump 1 and pump 2 come out to be $\SI{5.15}{\mega\hertz}$ and $\SI{4.26}{\mega\hertz}$ respectively. This results in $g_1/2\pi=\SI{0.53}{\mega\hertz}$ and $g_2/2\pi=\SI{0.48}{\mega\hertz}$.

\subsection{Wigner tomography}

The Wigner tomography of the storage resonator is performed in a similar manner to \cite{Lutterbach1997,Sun2014}. After preparing the storage resonator in the desired state, we apply a displacement pulse on the storage resonator, displacing it by $\beta$. Following this, we perform a non-demolition measurement of the parity of the storage resonator using the tomography transmon. A narrow unselective gaussian pulse ($\sigma=\SI{20}{\nano\second}$)  puts the transmon in the superposition of ground and excited state irrespective of the state of the storage resonator. Next, the transmon undergoes free evolution under the dispersive coupling with the resonator. By choosing the evolution time to be $\pi/\chi_{\rm transmon, resonator}=\SI{416}{\nano\second}$ and performing another $\pi/2$ pulse on the transmon, we map the parity of the resonator on the state of the transmon. Fig. 4 shows the average parity of the storage resonator as a function of the real and imaginary part of the applied displacement $\beta$ which is the Wigner function of the storage resonator.

\section{Simulations}
\label{sec:sim_details}

In this section we give the details of the simulation results presented in Fig. 3. The results are obtained by simulating the Lindblad master equation given by
\begin{align}
\dot\rho = -\frac{i}{\hbar}\left[H_{\rm sys},\rho\right]+\kappa_a \mathcal{D}\left[\bfa\right]\rho + \kappa_b \mathcal{D}\left[\bfb\right]\rho +\Gamma_{\phi,a} \mathcal{D}\left[\bfa^\dagger \bfa \right]\rho + \Gamma_{\phi,b} \mathcal{D}\left[\bfb^\dagger \bfb \right]\rho.
\end{align}
Here $H_{\rm sys}$ is the Hamiltonian of the system as quoted in Eq.~\ref{eq:H_sys}. We also take into account the relaxation rates $\kappa_{a/b}=\frac{1}{T_{1,a/b}}$ and the dephasing of the modes $\Gamma_{\phi,a/b}=\frac{1}{2T_{1,a/b}} + \frac{1}{T_{2,a/b}}$. The magnitude of all the quantities appearing in the Hamiltonian has been quoted in the main text except for the value of the global frequency shift $\delta$ which is found by using Eq.~\ref{eq:delta}. From the resulting density matrix we find the populations of the various Fock states by tracing out the transmon ($b$) and also the $\ket{g}$, $\ket{e}$, $\ket{f}$ state populations of the transmon by tracing out the resonator (a). \textcolor{\reviewcolor}{The results of numerical simulations are scaled such that the maximum and minimum of each trace matches with the maximum and minimum of the corresponding experimental data. This helps in accounting for the measurement contrast of the experimental data.} The results of this simulations are plotted in Fig.~3 and they compare well with the experimental results.

\bibliography{prospectus_citations}

\begin{figure}[t!]
\centering
\includegraphics[width=\textwidth]{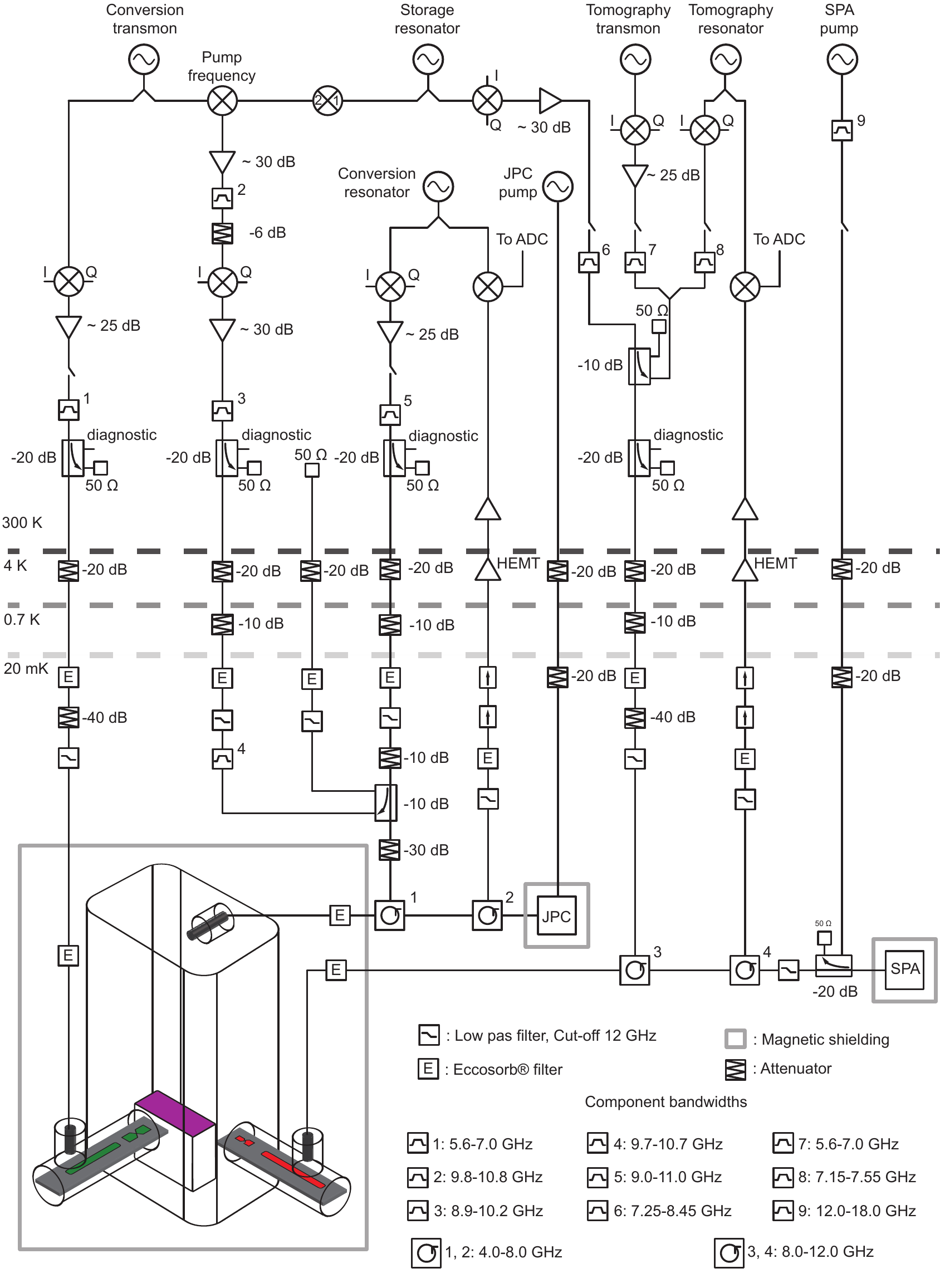}
\caption{\textbf{Wiring diagram.} }
\end{figure}